\documentclass{iau}
\usepackage{graphicx}

\title[Bayesian model comparison]
  {Bayesian model comparison in cosmology}

\author[D.\ J.\ Mortlock]
  {Daniel J.\ Mortlock}

\affiliation{Astrophysics Group, Blackett Laboratory,
and 
Department of Mathematics
\\ Imperial College London, London SW7 2AZ, United Kingdom \\
email: {\tt mortlock@ic.ac.uk}}

\pubyear{2014}
\volume{306}  
\pagerange{1--4}
\jname{Statistical Challenges in 21st Century Cosmology}
\editors{A.\ H.\ Heavens \& J.-L.\ Starck, eds.}
\begin{document}

\maketitle

\begin{abstract}
The standard Bayesian model formalism comparison cannot be applied 
to most cosmological models as they lack well-motivated parameter priors.
However, if the data-set being used
is separable then it is possible to use some of the 
data to obtain the necessary parameter distributions, the rest of the 
data being retained for model comparison. 
While such methods are not fully prescriptive, they provide a route
to applying Bayesian model comparison in cosmological situations where it could
not otherwise be used.
\end{abstract}


\newcommand{\vect}[1]{{\boldmath{#1}}}
\newcommand{\omm}{\Omega_{\rm m}}
\newcommand{\omlam}{\Omega_\Lambda}
\newcommand{\omlambb}{\Omega_{\Lambda,{\rm BB}}(\omm)}
\newcommand{\ommax}{\Omega_{\rm max}}
\newcommand{\model}{M}
\newcommand{\prob}{{\textrm{Pr}}}
\newcommand{\context}{I}
\newcommand{\data}{d}
\newcommand{\ie}{{\textit{i.e.}}}
\renewcommand{\eg}{\textit{e.g.}}
\newcommand{\cf}{\textit{c.f.}}
\newcommand{\redshift}{z}
\newcommand{\param}{\theta}
\newcommand{\params}{\{\theta_i\}}
\newcommand{\paramsprime}{\{\theta^\prime_i\}}
\newcommand{\diff}{{\rm{d}}}
\renewcommand{\etc}{{\em etc}.}
\newcommand{\accel}{{\rm accel.}}


\firstsection 
\section{Introduction}

Much of 
observational cosmology can be thought of 
as an attempt
to use astronomical data to 
discriminate between the different cosmological models
under consideration.
Given both the inevitably imperfect data and the intrinsically stochastic
nature of many cosmological measurements (\ie, cosmic variance),
it is generally impossible to come to absolute conclusions about 
the various candidate models; 
the best that can be hoped for is to evaluate the probabilities,
conditional on the the available data,
that each of the candidate models is the correct description of the
Universe.
The fact that there is, 
as far as is known,
just a single observable Universe
(\ie, there is no ensemble from which it has been drawn),
means that
such probabilities cannot be frequency-based, and must instead
must represent a degree of implication.
Self-consistency arguments then require 
(\cite[Cox 1946]{Cox:1946})
that these probabilities 
be manipulated and inverted using Bayes's theorem.

%

Taken together, the above facts imply that Bayesian 
model comparison (Section~\ref{section:bmc}) should be used to assess
how well different cosmological models
explain the available data, 
although the fact that most such models have unspecified parameters
is a significant difficulty for this approach
(Section~\ref{section:noprior}). 
This problem can be solved for separable data-sets 
as it is possible to use a two-step method of model comparison
(Section~\ref{section:sep}),
illustrated here with
high-redshift supernova (SN) data 
(Section~\ref{section:sn}).




\section{Bayesian model comparison}
\label{section:bmc}

Given that one of a set of $N$ models,
$\{\model_1, \model_2, \ldots \model_N\}$,
is assumed to be true, the state of knowledge conditional on 
all the available (and relevant) information, $\context$,
is fully summarised by the probabilities 
$\prob(\model_1 | \context), 
\prob(\model_2 | \context), 
\ldots,
\prob(\model_N | \context)$, 
where 
$\prob(\model_i | \context)$ is the 
probability that the $i$'th model is correct 
(and $i \in \{ 1, 2, \ldots, N\}$).
In the light of some new data, $\data$, that has not already
been included in the above probabilities, 
Bayes's theorem gives the updated probability that
model $i$ is correct
as 
\begin{equation}
\label{equation:bt}
\prob(\model_i | \data, \context)
=
\frac
{\prob(\model_i | \context) \, \prob(\data | \model_i, \context)}
{\sum_{j = 1}^N
\prob(\model_j | \context) \, \prob(\data | \model_j, \context)},
\end{equation}
where
$\prob(\data | \model_i , \context)$ is the 
marginal likelihood
under model $M_i$.

If model $M_i$ has $N_i$ unspecified parameters 
$\params = \{\param_{i,1}, \param_{i,2}, \ldots, \param_{i,N_i}\}$
then the model-averaged likelihood is obtained by marginalising over 
these parameters to give
\begin{equation}
\label{equation:ev}
\prob(\data | \model_i, \context)
= 
\int 
\prob(\params | \model_i, \context) \,
\prob(\data | \params, \model_i, \context) \,
\diff \param_{i,1} \, \diff \param_{i,2} \ldots \diff \param_{i,N_i},
\end{equation}
where $\prob(\params | \model_i, \context)$ is the 
prior distribution of the parameter values in this model.
This expression demonstrates that the 
full specification of a model 
requires not just an explicit parameterisation,
but a distribution for those parameters as well;
two mathematically identical descriptions with different parameter priors 
are, in fact, different models.


\section{Comparison of models without parameter priors}
\label{section:noprior}

Equations \ref{equation:bt} and \ref{equation:ev} 
together summarise a self-consistent method
for assessing which of a set of models is better supported by the 
available information, provided
that the parameter priors for all the models are explicitly defined
and unit-normalised.
In particular, while it is often possible to obtain sensible 
parameter constraints based on an improper prior, such as
$\prob(\params | \model_i, \context)$ constant for all $\params$,
the resultant marginal likelihood is meaningless
(\cite[Dickey 1961]{Dickey:1961}).
Unfortunately, it is commonly the case in astronomy and cosmology
that there is no compelling form for the models' parameter priors
and, further, that the natural uninformative prior distributions are
improper and cannot be normalised.
The apparent implication is
that Bayesian model comparison,
at least in the form described in Section~\ref{section:bmc},
cannot be used in cosmology,
an idea that has been explored previously by,
\eg, \cite{Efstathiou:2008} 
and 
\cite{Jenkins_Peacock:2011}.
The disturbing corollary would be that 
there is no self-consistent method to 
choose between the available cosmological models,
even if they are 
completely 
quantitative and mathematically well-defined.


\section{Model comparison with separable data}
\label{section:sep}

The idea that the relative 
degree of support for models with unspecified parameters is undefined 
is at odds with the marked -- and data-driven -- progress that has
been made in cosmology over the last century.
%
Clearly it {\em is} possible to use data to choose sensibly
between models even if they do not have well-motivated parameter priors; 
but can this be formalised in a way that satisfies Bayes's theorem
and is hence logically self-consistent?

One possibility is, for separable data-sets 
(such as those which consist of measurements of 
many astronomical sources),
to use some of the available data to obtain
the necessary parameter priors and to then use the remaining data
for model comparison.  
This is an old concept, dating back at least to \cite{Lempers:1971}
and explored subsequently by, \eg,
\cite{Spiegelhalter_Smith:1982}
and
\cite{OHagan:1995}.
The central idea is to partition the data as
$\data = (\data_1, \data_2)$, with the first partition 
of training data used to 
obtain the (partial) posterior distribution for the parameters of $i$'th model
as
\begin{equation}
\prob(\params | \data_1, \model_i, \context) 
  = \frac{\prob(\params | \model_i, \context) \, 
  \prob(\data_1 | \params, \model_i, \context)}
  {\int \prob(\paramsprime | \model_i, \context) \, 
  \prob(\data_1 | \paramsprime, \model_i, \context) \, 
\diff \param_{i,1}^\prime \, \diff \param_{i,2}^\prime 
  \ldots \diff \param_{i,N_i}^\prime,
},
\end{equation}
where $\prob(\params | \model_i, \context)$,
which need {\em not} be normaliseable,
should be a highly uninformative prior.
This posterior distribution can then be used as the prior needed
to obtain a meaningful marginal likelihood, which can then be evaluated
for the testing data as
\begin{equation}
\prob(\data_2 | \data_1, \model_i, \context) 
  = \int \prob(\params | \data_1, \model_i, \context) 
  \, \prob(\data_2 | \params, \model_i, \context) 
 \, \diff \param_{i,1} \, \diff \param_{i,2} \ldots \diff \param_{i,N_i}
  .
\end{equation}

This marginal likelihood is coherent, in the sense that it provides
self-consistent updated posterior probabilities when inserted into
Equation~\ref{equation:bt}, but there is also ambiguity in how to
partition the data:
there is no compelling scheme for
partitioning the data.
It is tempting to average over the possible
partitions, but this approach does not have a rigorous motivation.
Despite these ambiguities, this two-step method of 
Bayesian model comparison for separable data does 
satisfy the \cite[Cox (1946)]{Cox:1946} self-consistency
requirements and so 
provide a means 
of calculating posterior probabilities for cosmological
models with unspecified parameter priors.



\section{Example: late-time acceleration and supernovae}
\label{section:sn}

\begin{figure}[t]
\begin{center}
\includegraphics[width=6cm]{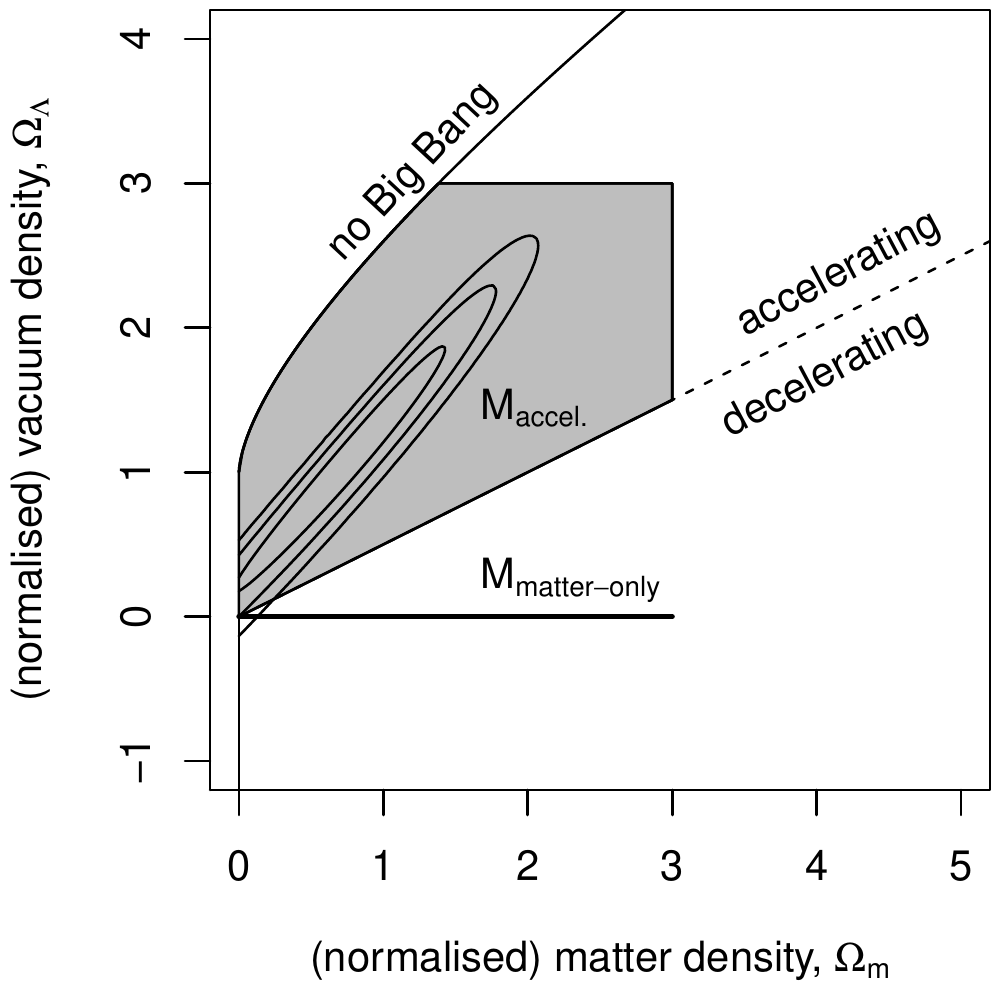}
\includegraphics[width=6cm]{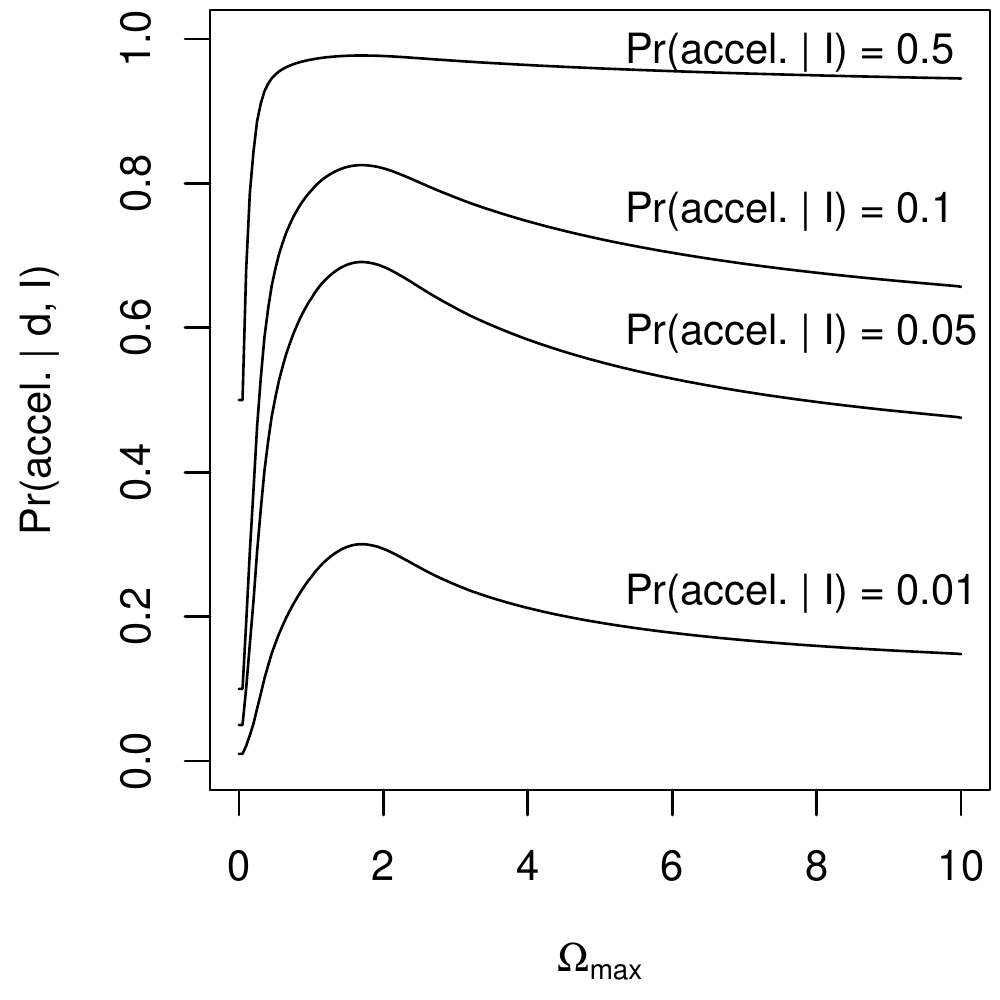}
\caption{(left) The posterior distribution of $\omm$ and $\omlam$
  implied by the \cite{Perlmutter_etal:1999} SCP SN data
  and a uniform prior with $\omm \geq 0$.
  Highest posterior density contours enclosing 68.3\%, 95.4\%
  and 99.7\% of the posterior probability are shown.
  Also shown are the prior distributions of 
  the accelerating model and matter only model for $\ommax = 3$.
  (right) The dependence of $\prob(\accel |\data, \context)$
  on $\ommax$, shown for different prior probabilities, 
  $\prob(\accel | \context)$.}
\label{figure:posterior}
\end{center}
\end{figure}

One of the most significant recent cosmological discoveries was that
the Universe's expansion rate is increasing, 
a result which is often linked most strongly to 
the observations of distant SNe
made by \cite{Riess_etal:1998} and \cite{Perlmutter_etal:1999}.
The comparative faintness of the SNe,
given their redshifts and light-curve decay timescales,
indicated
that the (normalised) cosmological constant, $\omlam$,
is sufficiently large to override the deceleration 
caused by the (normalised) matter density, $\omm$.
\cite{Riess_etal:1998} and \cite{Perlmutter_etal:1999}
used their SNe measurements, $\data$,
to obtain posterior distributions of the form
$\prob(\omlam, \omm | \data, \context)$, 
under the assumption of 
uninformative (and improper) uniform 
priors of the form $\prob(\omm, \omlam) \propto \Theta(\omm)$,
where $\Theta(x)$ is the Heaviside step function.
The posterior distribution for the 42 SCP SNe from 
\cite{Perlmutter_etal:1999}, reproduced in 
Figure~\ref{figure:posterior}, reveals that 
most of the models that are consistent with the data correspond
to an accelerating universe (\ie, $\omlam > \omm / 2$).

But do these data provide {\em quantitive} evidence of
cosmological acceleration?  
\cite{Riess_etal:1998} approached this question by calculating the 
fraction of the posterior with $\omlam > \omm / 2$,
which is an apparently compelling 0.997 for the case shown 
in Figure~\ref{figure:posterior}.
The relevant Bayesian calculation 
(\cf\ \cite[Drell \etal\ 2000]{Drell_etal:2000})
should, however, 
be based on the marginal likelihoods of an
accelerating model (for which the prior is
non-zero only for $\omlam > \omm / 2$)
and a decelerating model
(for which the obvious option is a matter-only model with $\omlam = 0$).
Such models can be fully specified (in the sense defined in 
Section~\ref{section:bmc}) by adding the restrictions 
that $0 \leq \omm \leq \ommax$
and $0 \leq \omlam \leq \min[\ommax, \omlambb]$ 
(defined to reject models that did not begin with 
a Big Bang), where $\ommax \geq 0$ is an unspecified ``hyper-parameter''.
Figure~\ref{figure:posterior} shows the dependence of 
the posterior probability of the accelerating model,
$\prob(\accel | \data, \context)$,
on $\ommax$.  
Even the peak values of $\prob(\accel | \data, \context)$ are
considerably lower than the posterior
fraction quoted above,
and the dependence on the unknown value of $\ommax$ is significant as well.

\begin{figure}[t]
\begin{center}
\includegraphics[width=6cm]{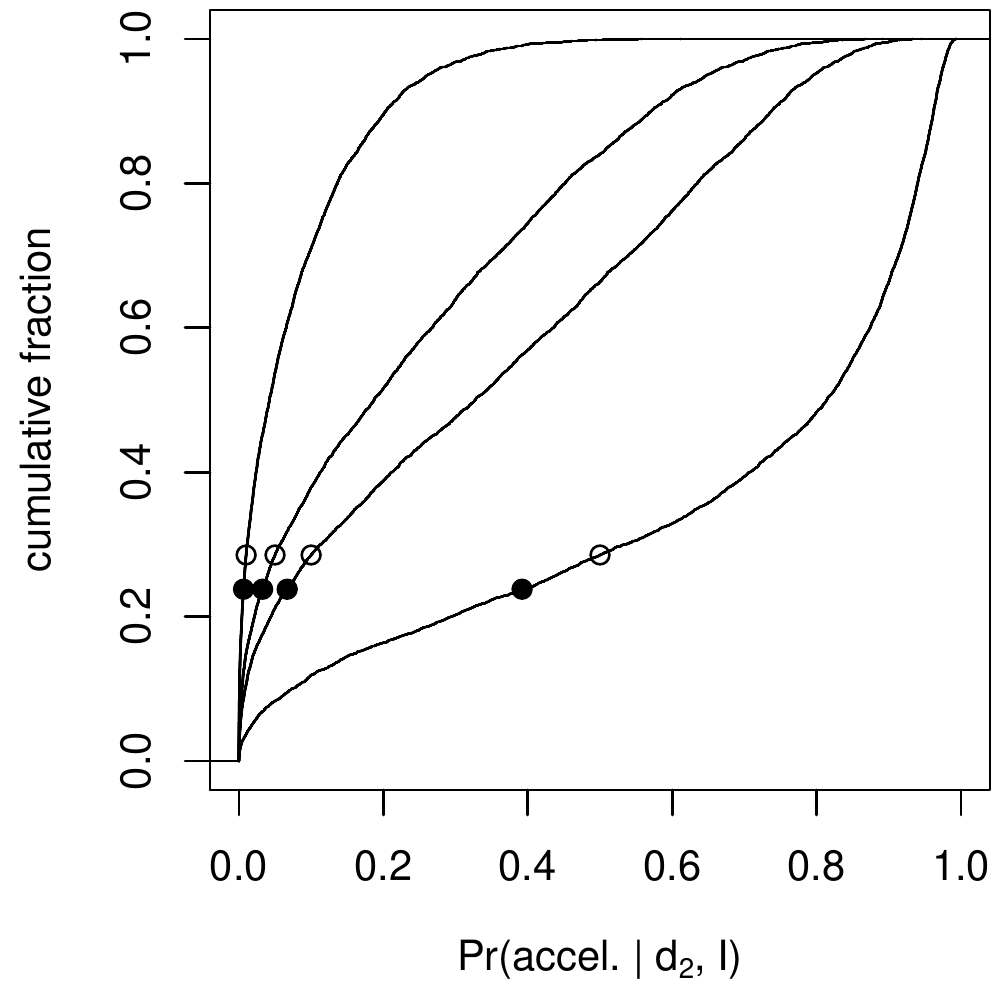}
\includegraphics[width=6cm]{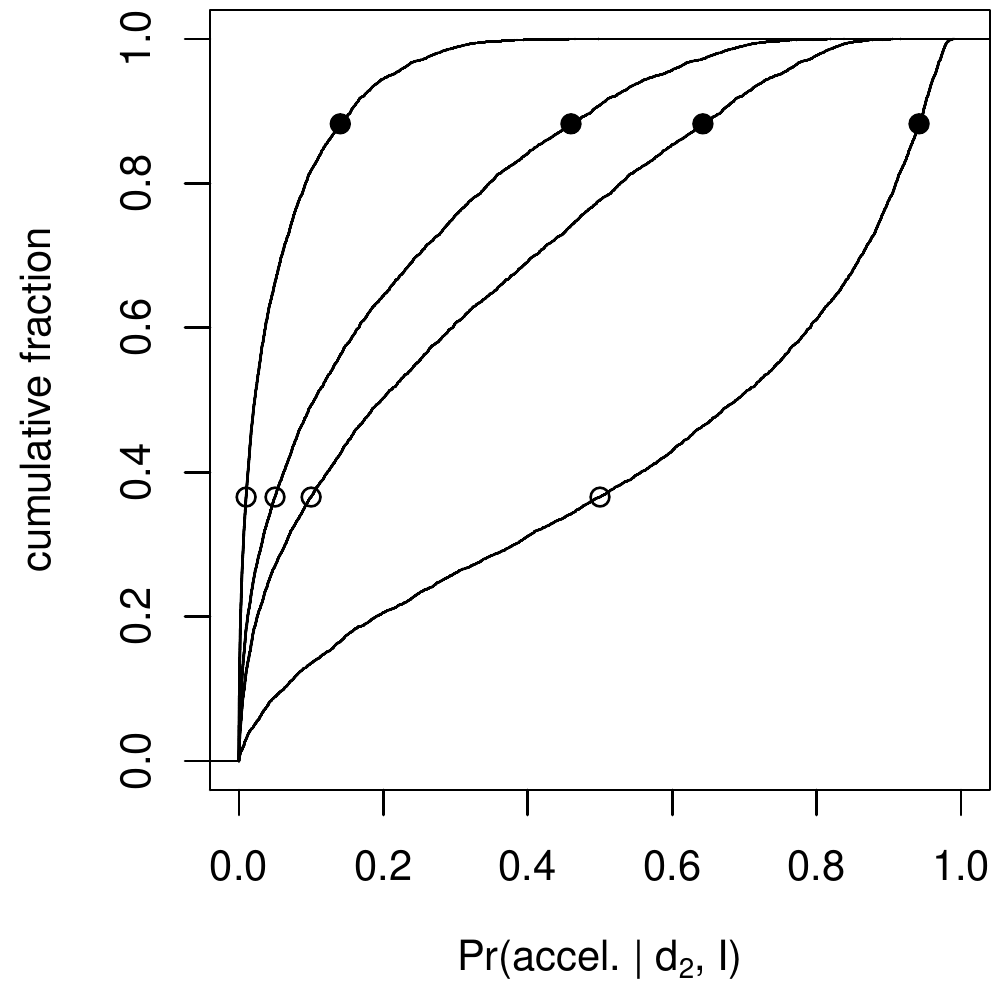}
\caption{The distribution of $\prob(\accel | \data_2, \context)$
obtained from different partitions of the 
\cite{Perlmutter_etal:1999} SN data set with training sets of 
10 (left) and 21 (right) SNe.
The open symbols indicate the prior values 
(of, from left to right, 0.01, 0.05, 0.1 and 0.5)
and the solid symbols show the posterior values given by 
training and testing samples that alternate in redshift.}
\label{figure:prob}
\end{center}
\end{figure}

Rather than introducing an arbitrary new parameter, another option 
is to adopt the two-step method described in Section~\ref{section:sep},
using some of the SN data to obtain a partial posterior in 
$\omm$ and $\omlam$ for both the accelerating and matter-only models
and then using the remainder to perform model comparison.
The results of doing so are shown in Figure~\ref{figure:prob}
for several different partitioning options
(and assuming the two models are equally probable a priori).
These results again illustrate the standard Bayesian result that 
the better-fitting accelerating model is not favoured so decisively 
over the more predictive (\ie, ``simpler'') matter-only model,
a result that is robust to prior choice.

This two-step approach to model comparison could be applied to a 
variety of problems in astrophysics and cosmology 
(\eg, \cite[Bailer-Jones 2012]{Bailer-Jones:2012},
\cite[Khanin \& Mortlock 2014]{Khanin:2014}).

%
%
%


%


\begin{thebibliography}{}

\bibitem[Bailer-Jones (2012)]{Bailer-Jones:2012}
  {Bailer-Jones, C.} 2012
  \textit{Astronomy \& Astrophysics}, 546, A89

\bibitem[Cox (1946)]{Cox:1946}
  {Cox, R.~T.} 1946, 
  \textit{American Journal of Physics}, 14, 1

\bibitem[Drell, Loredo \& Wasserman (2000)]{Drell_etal:2000}
  {Drell, P.~S., Loredo, T.~J.\ \& Wasserman, I.} 2000,
  \textit{The Astrophysical Journal},
  530, 593

\bibitem[Efstathiou (2008)]{Efstathiou:2008}
  {Efstathiou, G.} 2008,
  \textit{Monthly Notices of the Royal Astronomical Society}, 
  388, 1314


\bibitem[Jenkins \& Peacock (2011)]{Jenkins_Peacock:2011}
  {Jenkins, C.~R.\ \& Peacock, J.~A.} 2011,
  \textit{Monthly Notices of the Royal Astronomical Society}, 
  413, 2895

\bibitem[Khanin \& Mortlock (2014)]{Khanin:2014}
  {Khanin, A., Mortlock, D.\ J.} 2014,
  \textit{Monthly Notices of the Royal Astronomical Society},
  444, 1591

\bibitem[Lempers (1971)]{Lempers:1971}
  {Lempers, F.~B.} 1971,
  \textit{Posterior Probabilities of Alternative Linear Models},
  Rotterdam: University Press

\bibitem[O'Hagan (1995)]{OHagan:1995}
  {O'Hagan, A.} 1995, 
  \textit{Journal of the Royal Statistical Society B}, 30, 490 

\bibitem[Perlmutter \etal\ (1999)]{Perlmutter_etal:1999}
  {Perlmutter, S., \etal} 1999,
  \textit{The Astrophysical Journal},
  517, 565

\bibitem[Spiegelhalter \& Smith (1982)]{Spiegelhalter_Smith:1982}
  Spiegelhalter, D.~J.\ \& Smith, A.~F.~M., 1982,
  \textit{Journal of the Royal Statistical Society B}, 44, 377

\bibitem[Riess \etal\ (1998)]{Riess_etal:1998}
  {Riess, A.~G., \etal} 1998,
  \textit{The Astronomical Journal},
  116, 1009

\end{thebibliography}
\end{document}